\documentclass[11pt,a4paper]{article}

\usepackage[utf8]{inputenc}
\usepackage{amsmath}
\usepackage[mathcal]{euscript}
\usepackage{latexsym}
\usepackage{textcomp}
\usepackage{slashed}
\usepackage{tikz}
\usepackage{xspace}
\usepackage{setspace}
\usepackage[a4paper]{geometry}
\usepackage[numbers,sort&compress]{natbib}

\newcommand{\refeq}[1]{(\ref{#1})}

\newcommand{\refcite}[1]{Ref.~\cite{#1}}

\newcommand{\Dd}[0]{\frac{\Delta}{2}}

\newcommand{\ie}{\textit{i.e. }}
\newcommand{\eg}{\textit{e.g. }}

\newcommand{\ds}{Dyson-Schwinger\xspace}

\newcommand{\bs}{Bethe-Salpeter\xspace}

\newcommand{\DDF}[0]{\ensuremath{F^q(\beta,\alpha,t)}}
\newcommand{\DDG}[0]{\ensuremath{G^q(\beta,\alpha,t)}}

\begin{document}

\begin{center}
\begin{spacing}{1.5}

{\Large\bf Generalised Parton Distributions: A\\ \ds approach for the pion.} \\
\end{spacing}
\vskip 10mm

C.~M\scshape {ezrag}\footnote{cedric.mezrag@cea.fr}
\\[1em]
{\small {\it IRFU/Service de Physique Nucl\'eaire \\ CEA Saclay, F-91191 Gif-sur-Yvette, France}}\\

\end{center}

\begin{abstract}
\noindent  We compute the pion quark Generalised Parton Distribution H and quark Double Distributions in a coupled Bethe-Salpeter and Dyson-Schwinger approach in terms of quark flavors or isospin states. We use analytic expressions inspired by the numerical resolution of Dyson-Schwinger and Bethe-Salpeter equations. We obtain an analytic expression for the pion Generalised Parton Distribution at a low scale. Our model fulfils most of the required symmetry properties and compares very well to experimental pion form factor or valence parton distribution function experimental data. In addition, we have highlighted limitations of the so-called impulse approximation, which breaks symmetries when computing the valence parton distribution function. Doing so, we introduced new terms which were neglected before. Finally, we also shed light on a specific property of the pion GPD: Polyakov soft pion theorem.
\end{abstract}

\section{Introduction}

Generalised Parton Distributions (GPDs) were introduced in the 1990  \cite{Mueller:1998fv,Ji:1996nm,Radyushkin:1997ki} and since then have been deeply studied both theoretically (see \eg the reviews \cite{Diehl:2003ny,Belitsky:2005qn,Guidal:2013rya}) and experimentally. New results from Jefferson Laboratory facilities (JLab) on Deep-Virtual Compton Scattering (DVCS) on a proton target have recently been presented \cite{Defurne:2015lna,Jo:2015ema}. Several phenomenological parametrisations have been developed in order to fit the proton GPDs \cite{Goloskokov:2005sd,Guidal:2004nd,Kumericki:2009uq,Mezrag:2013mya,Polyakov:2008xm,Goldstein:2010gu}. Yet until now, none of them has been fully derived from QCD dynamics only.

In order to generate the parton structure of hadron in a dynamical way, one can turn to the \ds (DS) and \bs (BS) equations. Those equations are coupled and can be solved using specific truncation schemes. In the following, we focus on the pion GPD, which is simpler to compute in the present framework than the proton one. The pion GPD inspired many theoretical studies \cite{Tiburzi:2002tq, Theussl:2002xp, Bissey:2003yr, VanDyck:2007jt, Frederico:2009fk,Dorokhov:2011ew} in spite of the restricted set of existing experimental data, related to the valence Parton Distribution Function (PDF) \cite{Aicher:2010cb}  and the form factors \cite{Amendolia:1986wj,Huber:2008id}. Our approach to the pion GPD is detailed in the first section. The second section deals with the restoration of the symmetry under the exchange $x \leftrightarrow 1-x$ in the forward case. 

\section{GPD Computations}

\begin{figure}[h]
  \centering
  \includegraphics[width=0.27\textwidth]{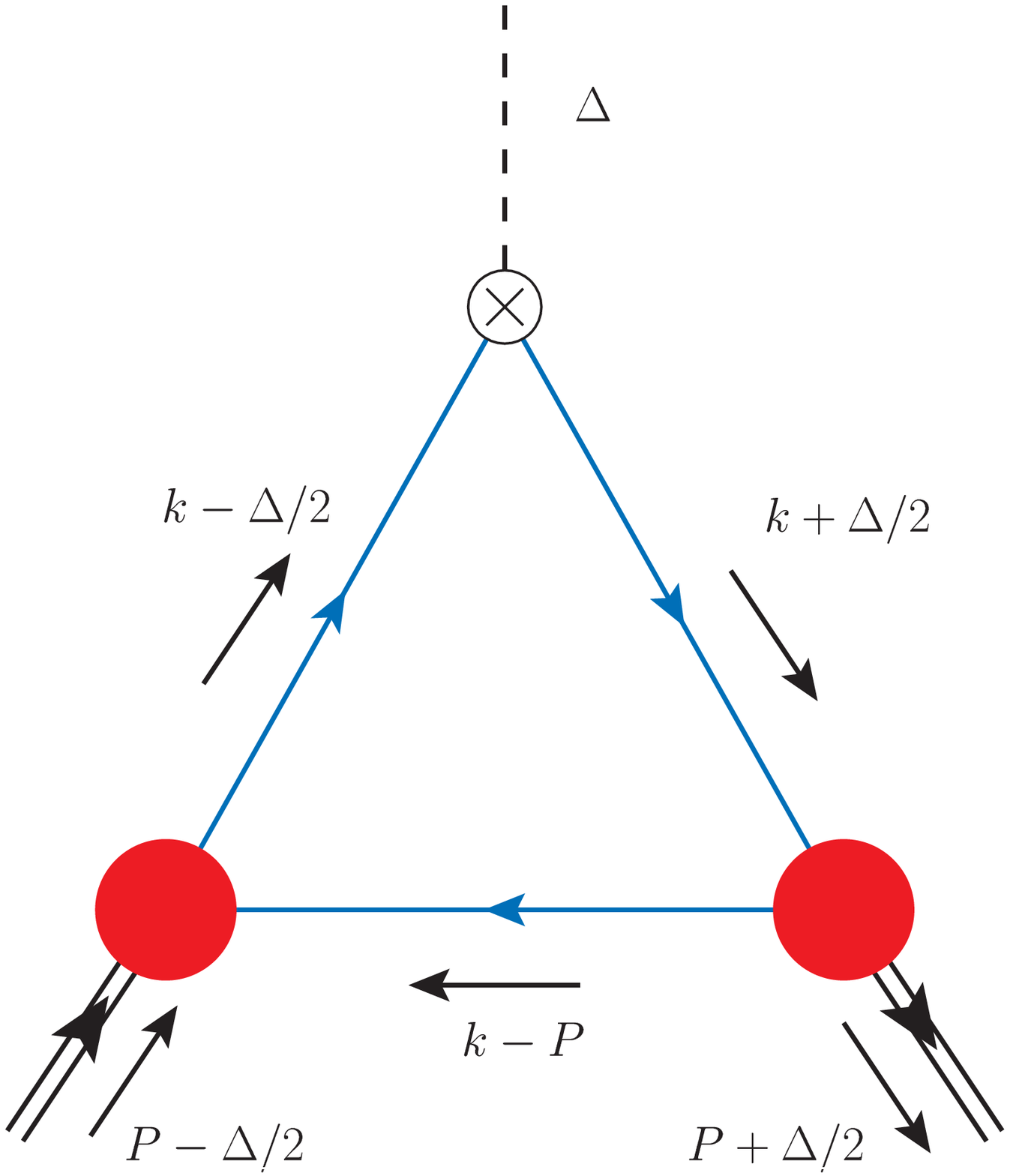}
  \quad
  \includegraphics[width=0.3\textwidth]{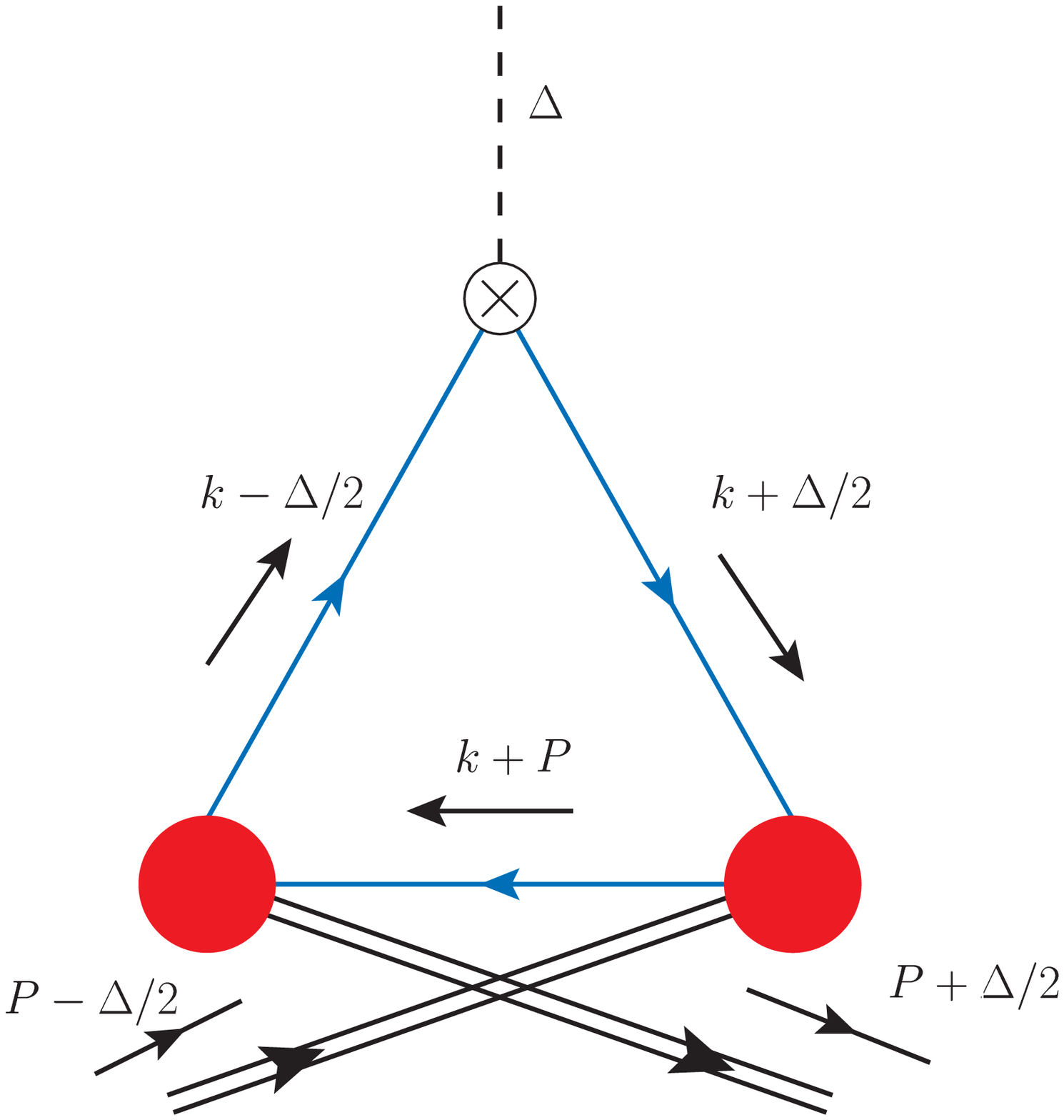}
  \caption{Mellin moments computed in the triangle approximation for the pion valence dressed quarks. \emph{Left panel}: quark case. \emph{Right Panel}: anti-quark case.}
  \label{fig:TriangleDiagram}
\end{figure}

\begin{figure}[h]
  \centering
  \includegraphics[width=0.3\textwidth]{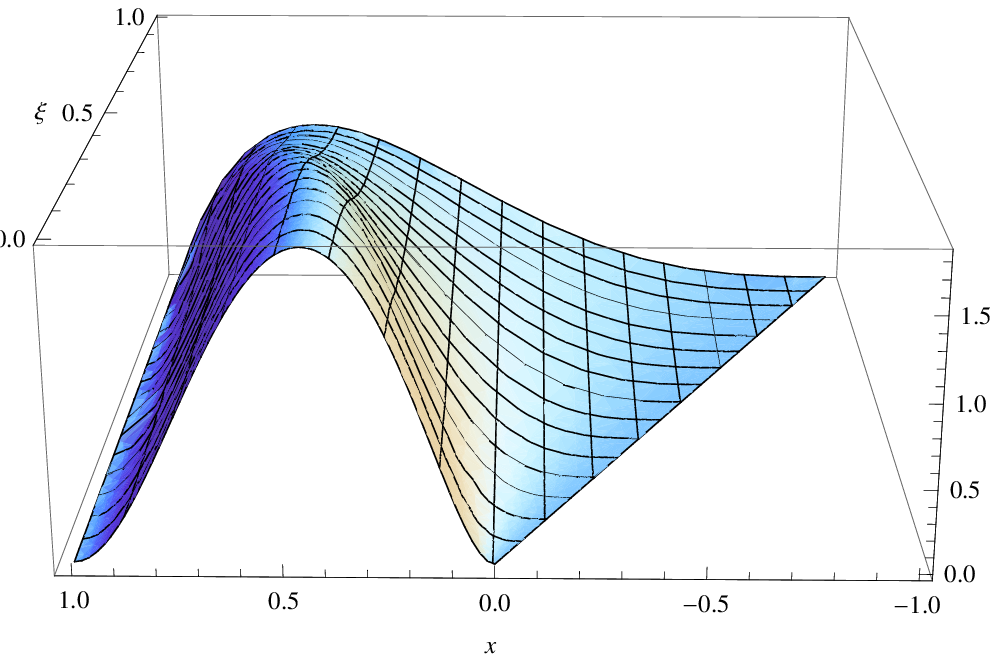}
  \quad
  \includegraphics[width=0.3\textwidth]{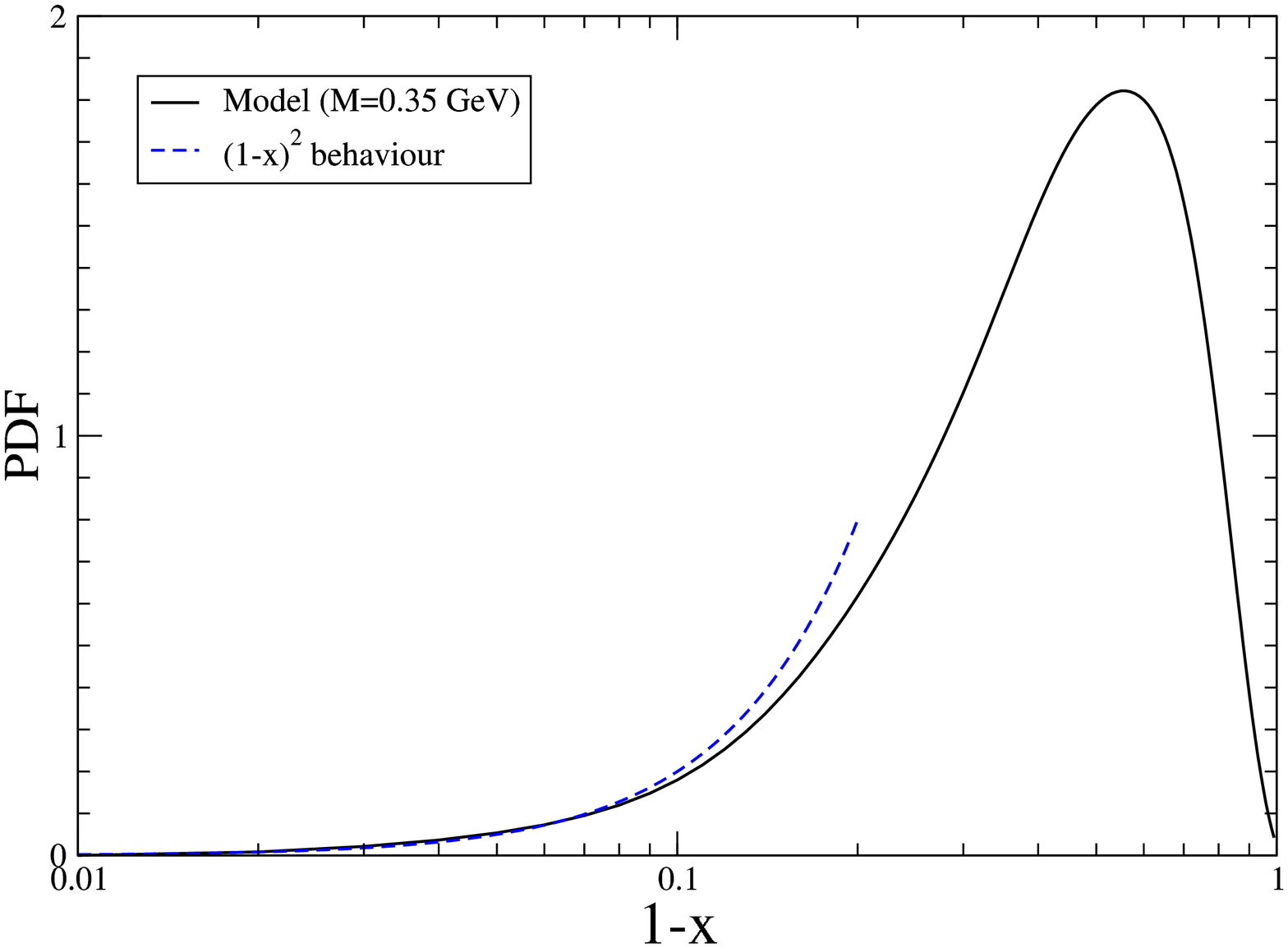}
  \caption{\emph{Left Panel}: reconstruction of the GPD when $t \rightarrow 0$. Support and continuity properties are well illustrated on this plot. \emph{Right Panel}: PDF large-$x$ behaviour, which fits the one predicted in the pertubative case.}
  \label{fig:GPD}
\end{figure}

The pion quark GPD $H^q(x,\xi,t)$ is formally defined as:
\begin{equation}
   \label{eq:DefinitionGPDH}
   H(x,\xi,t) = \frac{1}{2} \int \frac{\textrm{d}z^-}{2\pi} \, e^{i x P^+ z^-} \left\langle P+\frac{\Delta}{2} \left| \bar{q}\left(-\frac{z}{2}\right)\gamma^+q\left(\frac{z}{2}\right) \right |P-\frac{\Delta}{2}\right\rangle_{z^+=0,z_\perp=0},
\end{equation}
where $t = \Delta^2$ and $\xi = -\frac{\Delta ^+}{2 P^+}$. In our approach \cite{Mezrag:2014tva,Mezrag:2015xia}, we define the pion GPD by the set of its Mellin moments $\mathcal{M}_n(\xi,t)$:
\begin{equation}
  \label{eq:DefinitionMellinMoments}
  \mathcal{M}_n(\xi,t) = \int \textrm{d}x~x^nH(x,\xi,t).
\end{equation}
These moments are computed in the triangle diagram approximation illustrated on figure \ref{fig:TriangleDiagram}.
Consequently, the inserted operators, depicted on figure \ref{fig:TriangleDiagram} by a cross, are the local twist-two quark operators arising from the operator product expansion. They can be computed as:
 \begin{eqnarray}
  \label{eq:IntegralMellinMoments}
  2( P \cdot n )^{m+1} \mathcal{M}_m(\xi,t) & = & \textrm{Tr} \int \frac{\mathrm{d}^4k}{(2\pi)^4} \, ( k \cdot n )^m i\Gamma_\pi( k -\Dd, P - \Dd )~ S( k - \frac{\Delta}{2} ) \nonumber\\
         & & i\gamma \cdot n~ S( k+\frac{\Delta}{2} )~i\bar{\Gamma}_\pi( k+\Dd,  P+\frac{\Delta}{2} )~ S( k - P ),
\end{eqnarray}
where the \bs vertices $\Gamma_\pi$ and propagator $S$ are taken as:
\begin{eqnarray}
S( p ) 
& = & \big[ - i \gamma \cdot p + M \big] \Delta_M( p ^2 ), \label{eq:QuarkPropagator} \\
\Delta_M( s )
& = & \frac{1}{s + M^2}, \label{eq:PropagatorMassTerm} \\
\Gamma_\pi( k, p )
& = & i \gamma_5 \frac{M}{f_\pi}M^{2\nu} \int_{-1}^{+1} \mathrm{d}z \, \rho_\nu( z ) \ 
\left[\Delta_M( k_{+z}^2 )\right]^\nu; \label{eq:Vertex} \\
\rho_\nu( z ) & = & R_\nu ( 1 - z^2 )^\nu. \label{eq:RhoFunction} 
\end{eqnarray}
These functional forms have been introduced in \refcite{Chang:2013pq} and allow one to fit the numerical solutions of the \ds and \bs equations. Here we will fit a single free parameter, $M$, the second one, $\nu$, being fixed to 1. This value of $\nu$ is the one giving back the pion asymptotic Distribution Amplitude (DA). From that point, the GPD can be analytically reconstruct from its Mellin using the so-called Double Distributions (DDs). Indeed, performing the computation, it is possible to identify the two DDs \DDF~and \DDG~which are related to the pion GPD through the Radon transform:
\begin{equation}
  \label{eq:DD}
  H_\pi^q(x,\xi,t) = \int_{-1}^1 \textrm{d}\beta \int^{1-|\beta|}_{-1+|\beta|} \textrm{d}\alpha \Big(F^q(\beta,\alpha,t)+ G^q(\beta,\alpha,t)\Big)\delta(x-\beta-\xi \alpha ) .
\end{equation}
It is then possible to get an algebraic formula for the GPD. The result is plotted on figure \ref{fig:GPD}.
Being able to compute algebraically the GPD provides us a significant advantage to check the different intrinsic properties that the model should respect. Due to the property of the algebraic DDs identified from our calculations, it is possible to conclude that the support property of the valence GPD, the continuity on the lines $x = \pm \xi$ and the parity in $\xi$ are fulfilled by our GPD model. Moreover, the DD formalism ensure the polynomiality property by construction. Finally, one can check that the resulting PDF gets the same large-$x$ behaviour than the one predicted by perturbation theory.

\section{Soft Pion theorem}

In \refcite{Polyakov:1998ze}, it has been shown that, it was possible to relate the pion DA to the pion GPD. In the kinematic limit when $t\rightarrow 0$ and $\xi \rightarrow 1$, one actually gets:
\begin{equation}
  \label{eq:SoftPionTheorem}
  H^u(x,1,0) = \frac{1}{2} \varphi_\pi\left( \frac{1+x}{2} \right),
\end{equation}
where $\varphi_\pi$ is the pion DA.
The algebraic model coming from equations \refeq{eq:QuarkPropagator}-\refeq{eq:RhoFunction} have been developed in order to describe $\varphi_\pi$. And indeed, when computed in \refcite{Chang:2013pq}, it leads to the asymptotic pion DA. Therefore, one could expect that, due to the soft pion theorem, when $\xi$ goes to $1$ and $t$ vanishes, the algebraic GPD model tend to the pion asymptotic DA. It is not the case due to a specific feature of the algebraic model. Indeed, working only with the building blocks of the solution of the DS-BS equations leads to a violation of the Axial-Vector Ward-Takashi identity (AVWTI), which gives in the chiral limit:
\begin{equation}
  \label{eq:AVWTI}
  P^\mu\Gamma_{5\mu}^i(k,P) = \frac{\tau^i}{2}\left(S^{-1}(k-\frac{P}{2})i\gamma_5 + i\gamma_5 S^{-1}(k+\frac{P}{2})\right),
\end{equation}
where $P$ is the total momentum entering the axial-vector vertex, $k$ the relative one, and the $\tau^i$ the Pauli matrices. When computed consistently in the rainbow ladder truncation scheme, the full solution of the DS-BS equation fulfils the AVWTI. Therefore, inserting the reduction formula:
\begin{eqnarray}
  \label{eq:AVReduction}
  2 f_\pi \Gamma_\pi(k; p ) &\stackrel{p \simeq 0}{\propto}&
  p^{\mu} \Gamma_{5\mu}(k,p)\,, \\
  \label{eq:ScalarReduction}
  2r_\pi\Gamma_\pi(k;p) &\stackrel{p^2 \simeq 0}{\propto}&
  p^2\Gamma_{5}(k,p)\, ,
\end{eqnarray}
allows one to write the triangle diagram in terms of axial-vector and pseudo-scalar vertices. Then, as shown in \refcite{Mezrag:2014jka}, most of the gluon ladders coming from the insertion of the AVWTI compensate each others, leading to the pion DA. Consequently, in the rainbow ladder truncation scheme, the soft pion theorem is automatically fulfilled providing that the \ds and \bs equations are solved consistently with the AVWTI.

\section{Additionnal contributions}

\begin{figure}[h]
  \centering
  \includegraphics[width=0.35\textwidth]{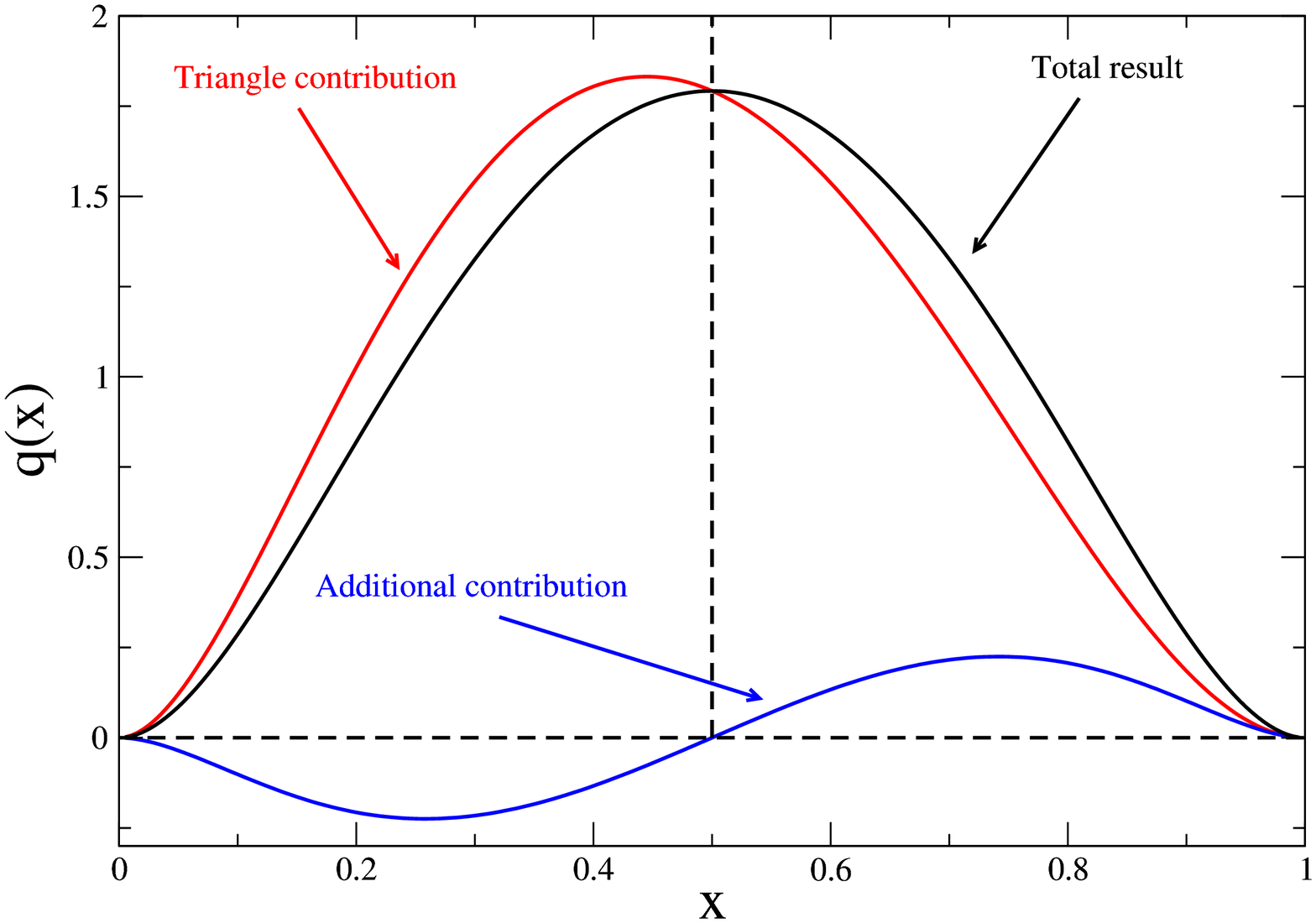}
  \caption{Contributions to the PDF and their sum.}
  \label{fig:PDFTotal}
\end{figure}

If it is possible to recover the soft pion theorem through a triangle diagram computation, such an approximation still has its own limitations. Focusing on the forward limit (\ie $t = 0$ and  $\xi = 0$), the PDF $q_\pi^{\textrm{Tr}}(x)$ can be written as:
\begin{align}
  \label{eq:PDFTriangle}
  q_\pi^{\textrm{Tr}}(x) = \frac{72}{25} \Big( ( 30 - 15 x + 8 x^2 - 2 x^3 )  x^3 \log x & + ( 3 + 2 x^2 ) ( 1 - x )^4 \log( 1 - x ) \\ \nonumber 
  & + ( 3 + 15 x + 5 x^2 - 2 x^3 ) x ( 1 - x ) \Big).
\end{align}
If this PDF is in agreement with the pertubative prediction at large $x$, it suffers a significant drawback, as it is not symmetric with respect to the $x \leftrightarrow 1-x$ exchange. The small asymmetry is due to the fact that contributions have been neglected, due to the triangle approximation. Indeed, it has been shown in \refcite{Chang:2014lva}, that one has to add new terms in order to fulfil the $x \leftrightarrow 1-x$ symmetry:
\begin{equation}
  \vcenter{\hbox{\includegraphics[width=0.15\textwidth]{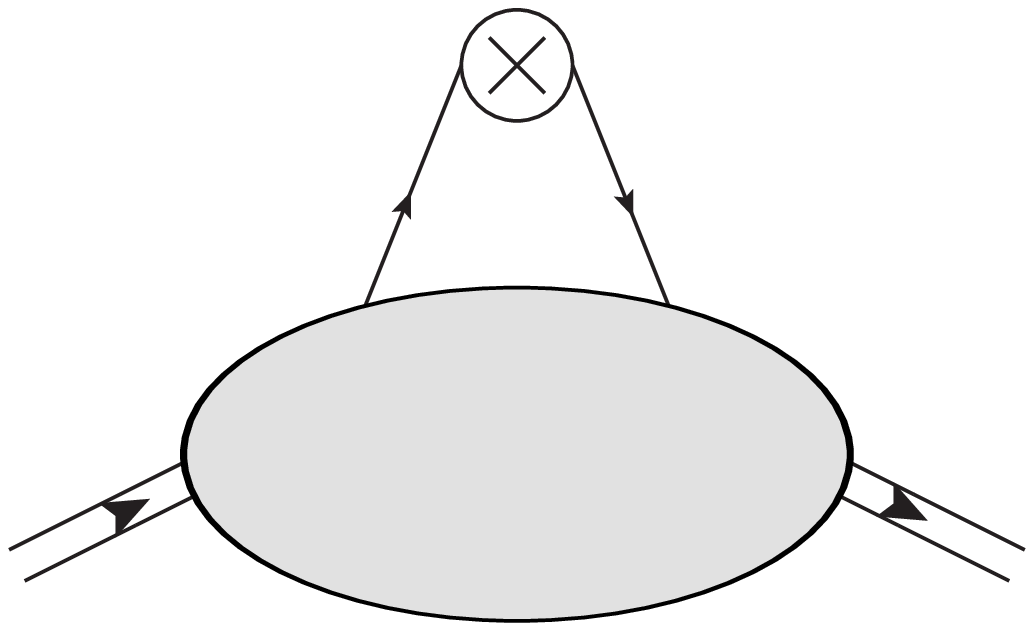}}} \simeq  \vcenter{\hbox{\includegraphics[width=0.15\textwidth]{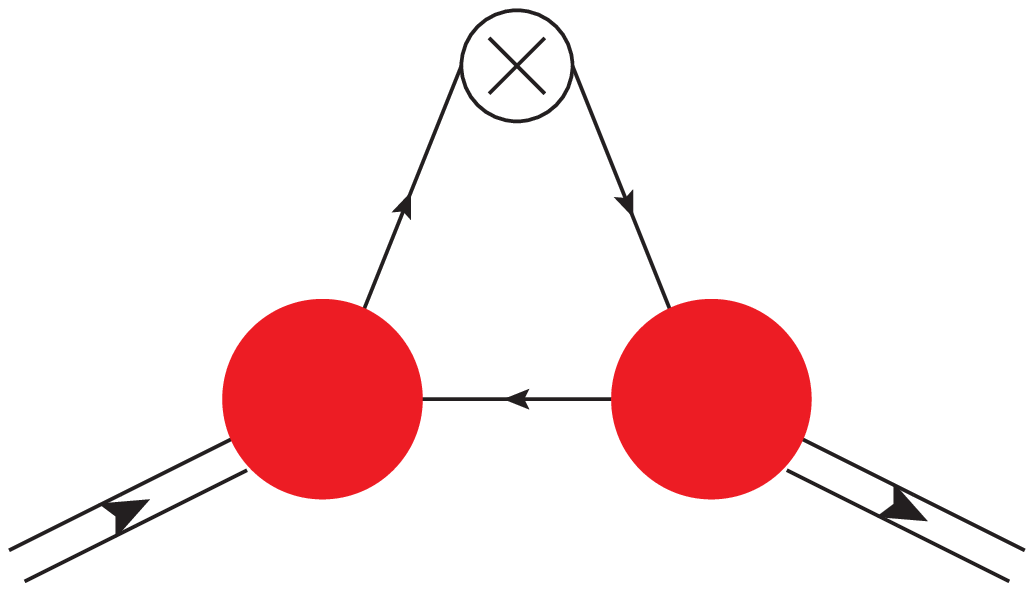}}} +  \vcenter{\hbox{\includegraphics[width=0.15\textwidth]{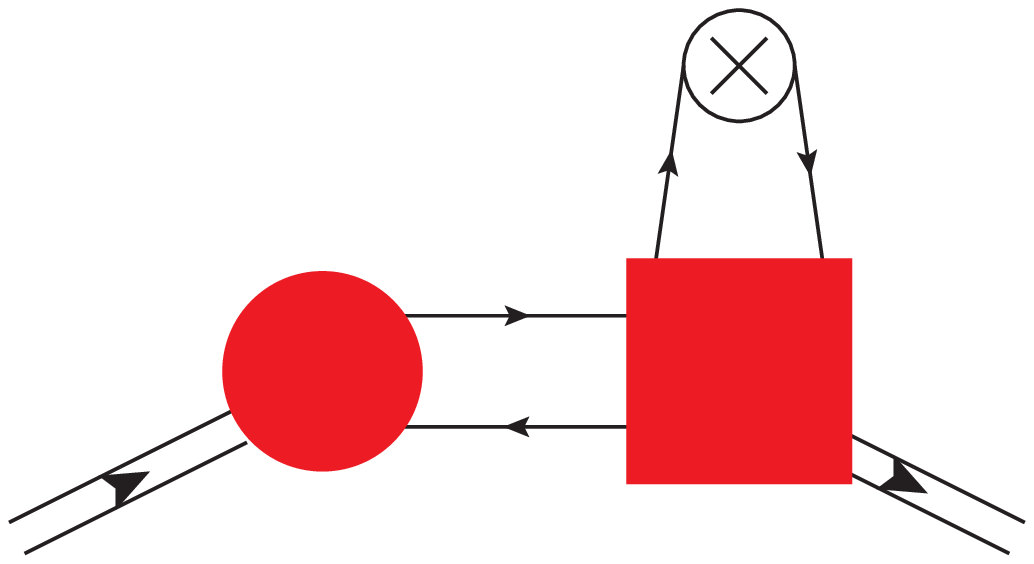}}} + \vcenter{\hbox{\includegraphics[width=0.15\textwidth]{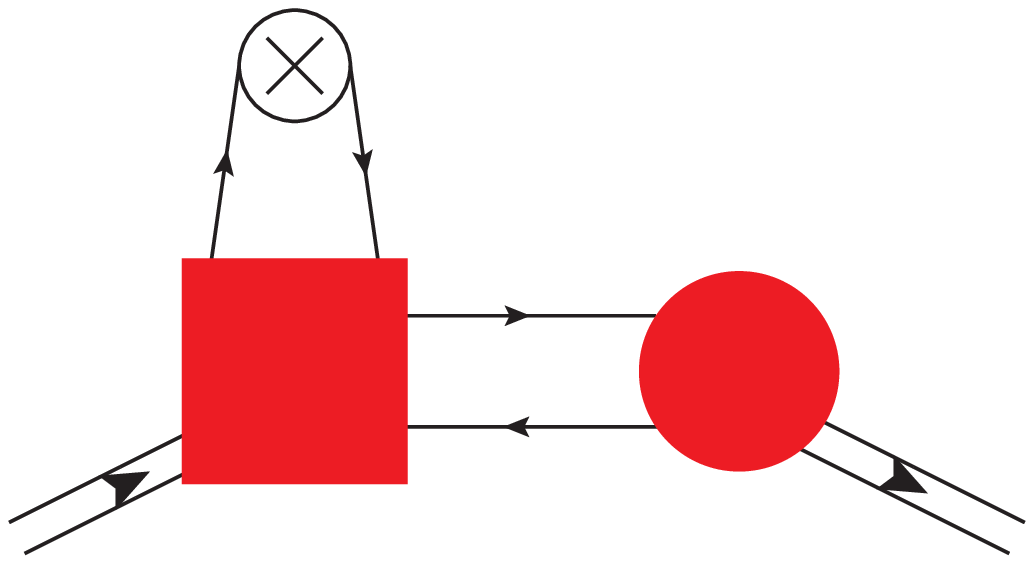}}} ,
\end{equation} 
where:
\begin{equation}
  \label{eq:FinalSquareVertex}
   \vcenter{\hbox{\includegraphics[width=0.20\textwidth]{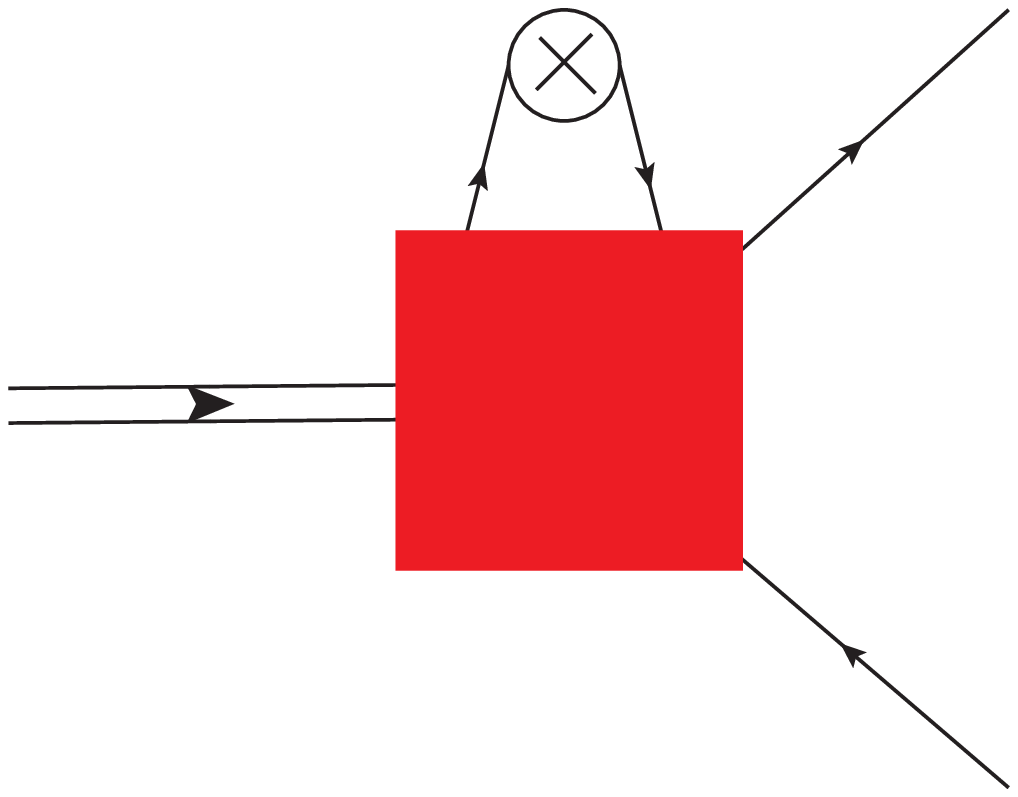}}} = -\frac{1}{2}(k \cdot n)^m n^\mu\frac{\partial \Gamma_\pi^q}{\partial k^\mu}\left(k, P \right).
\end{equation}
It is then possible to compute and additional component to the PDF $q_\pi^{\textrm{ad}}(x)$:
\begin{align}
   q_\pi^{\textrm{ad}}(x)  = \frac{72}{25} \Bigg(-\left(2 x^3+4 x+9\right) (x-1)^3 \log (1-x) & +x^3 (2 x ((x-3) x+5)-15) \log (x) \nonumber \\
   & - x (x-1)(2x-1) ((x-1) x-9) \Bigg) .
\end{align}
Then summing the different contributions, one gets the total PDF $q_\pi^{\textrm{tot}}(x)$:
\begin{eqnarray}
  \label{eq:PDFTotal}
  q_\pi^{\textrm{tot}}(x) & = & q_\pi^{\textrm{Tr}}(x) + q_\pi^{\textrm{ad}}(x) \nonumber \\
  & = & \frac{72}{25} \Bigg(x^3 (x (2 x-5)+15) \log (x)-\left(2 x^2+x+12\right) (x-1)^3 \log (1-x)\nonumber \\
  &  & -2 x  (x-1)((x-1) x+6)\Bigg),
\end{eqnarray}
which fulfils the $x \leftrightarrow 1-x $ symmetry as shown on figure \ref{fig:PDFTotal}.This analysis has also led to a new GPD Ansatz described in \refcite{Mezrag:2014jka,Chang:2015ela}.

\section{Conclusion}

We have developed a new GPD model. It is based on algebraic building blocks of solutions of the \ds and \bs equations. Within this procedure, our model fulfils most of the required theoretical constraints. In addition, our model is in good agreement with available experimental data. Computations has been done analytically up to the end, allowing one to identify limitations of the triangle diagram approximation, and thus to improve it. We have also emphasised the key-role of the AVWTI in the realisation of the soft pion theorem. Future models of pion GPDs will therefore have to deal with those additional constraints.

\section{Acknowledgement}

I thank my collaborators L. Chang, H. Moutarde, C.D. Roberts, J. Rodriguez-Quintero, F. Sabati\'e and P.C. Tandy with whom this work has been done. I also thank A. Besse, P. Fromholz, P. Kroll, J-Ph. Landsberg, C. Lorc\'e, B. Pire and S. Wallon for valuable discussions.  
This work is partly supported by the Commissariat \`a l'Energie Atomique, the Joint Research Activity "Study of Strongly Interacting Matter" (acronym HadronPhysics3, Grant Agreement n.283286) under the Seventh Framework Programme of the European Community, by the GDR 3034 PH-QCD "Chromodynamique Quantique et Physique des Hadrons", the ANR-12-MONU-0008-01 "PARTONS".

\bibliographystyle{hunsrt}
\bibliography{/home/cedric/Work/paper/Bibliography}

\end{document}